\documentclass[final,1p,times]{elsarticle} % do not change
\usepackage{graphicx} % do not change
\usepackage{amssymb} % do not change
\usepackage{amsthm} % do not change
\journal{Nuclear Physics A} % do not change
\begin{document} % do not change

\begin{frontmatter} % do not change

%% QM09Author: please enter your
%% Title, author and address info here; please do not use footnotes

% Your Title - please modify
\title{Photon Physics in ALICE}

% Principle author, and co-authors - please modify
\author{Yuri Kharlov$^{a}$ and Dmitri Peressounko$^{b}$ for the ALICE collaboration}

% Address - please modify
% note that if you have authors from several institutions, we recommend
% labelling these [a], [b], [c] etc.
\address[a]{IHEP, % label [a]
Pobeda str., 1, Protvino, 142281, Russia}
\address[b]{RRC "Kurchatov Institute", % label [b]
Kurchatov sq.,1, Moscow, 123128, Russia}

\begin{abstract} % do not change
We give an overview of photon physics which will be studied by the
ALICE experiment in proton-proton and heavy ion collisions at LHC.
We compare properties of ALICE photon detectors and estimate their
ability to measure neutral meson and direct photon spectra as well
as gamma-hadron and gamma-jet correlations in pp and Pb+Pb
collisions.
\end{abstract} % do not change

\end{frontmatter} % do not change

%% QM09: we keep linenumbers at least for initial version
%\linenumbers % do not change

\section{Physics of photons}

Almost any topic of the physics of heavy ion collisions can be
explored using photons. Decay photons can be used to measure
spectra, flow, nuclear modification factors and more complicated
observables of neutral mesons:  $\pi^0$, $\eta$ and $\omega$ mesons.
Direct photons provide possibility to explore the initial state of a
collision and evolution of the fireball through the thermal
emission. Separation of isolated and inclusive direct photons
provides the possibility to evaluate contributions of different
processes of direct photon emission and to look at modification of
jets in the medium. Finally constructing the photon-hadron and
photon-jet correlations one can produce parton fragmentation
functions and study their modifications in medium. All these topics
are well developed now and widely presented in this conference, see
e.g reviews of PHENIX and STAR results \cite{PHENIX-STAR}.

\section{Alice setup}

ALICE has an excellent possibility to explore all variety of signals in
heavy ion collisions. In addition to an advanced tracking system,
ALICE includes 2 electromagnetic calorimeters: PHOton Spectrometer
(PHOS) and EMCAL. The PHOS is a calorimeter with very fine
granularity and high energy and position resolutions. Its
primary goal is a precise measurement of the low and medium energy
$\pi^0$ and direct photons, however, it is able to measure photons
with energies up to 100 GeV. The second calorimeter, EMCAL, installed
back-to-back to PHOS, is concentrated on the measurement of hard
photons and jets. It has a larger acceptance but coarser granularity
and a moderate energy resolution. In addition to these
calorimeters, photons in ALICE can be measured via their conversion in
the central tracking system. This method has a rather low efficiency,
which is compensated to a large extend by a large acceptance of the
ALICE tracking system, so that
the product acceptance~$\times$~efficiency for the $\pi^0$ detection
is only 20 times smaller than that for the PHOS.

\begin{table}[ht]
\centering %%
\caption[]{Comparison of the properties of the ALICE photon
detectors.} \label{calo-comp}
\begin{tabular}{l|c|c|c}
  \hline
                   & PHOS  & EMCAL & conversion \\
  \hline
  $\sigma_{E}$/E (\%)  & $\sqrt{(1.3/E)^2+3.3^2/E+1.12^2}$ & $\sqrt{(11.3/E)^2+1.7^2/E+4.8^2}$ & 2 \\
  $\sigma_x$ (mm)      & $\sqrt{3.26^2/E+0.44}$ & $1.5+5.3/\sqrt{E}$ & $\sim 1$ \\
  $\sigma_{\pi}^{1<p_t<2 GeV/c}$ (MeV) & 5.5 & 16 & 3.3  \\
  Accept., $\Delta \phi \times \Delta \eta$
  &$100^{\circ}\times 0.25$ & $100^{\circ}\times 1.4$ & 2$\pi\times
  1.8$ \\
  \hline
\end{tabular}
\end{table}

Comparison of the energy and position resolutions of ALICE photon
detectors is presented in Table\ \ref{calo-comp}. The PHOS and the
central tracking system provide comparable energy and angular
resolutions somewhat better than those of the EMCAL. The best way to
compare resolutions of calorimeters is to look at the width of the
$\pi^0$ peak. We present the widths we expect at $1<p_t<2$ GeV/c in
pp collisions in Table\ \ref{calo-comp}. For the PHOS and the
conversion method the widths are comparable, while for the EMCAL the
width is $\sim 3$ times larger. So the ALICE incorporates 3 photon
detectors providing comparable resolutions. All three use different
experimental techniques which result in different systematic errors
and insure a good cross-check of the results.

\section{Measurement of the neutral meson and direct photon spectra}

There are 2 methods of detection of $\pi^0$ through its decay
$\pi^0\to 2\gamma$ in a calorimeter: at low $p_t$ the decay photons
are well separated and can be detected as two independent clusters so
that invariant mass analysis can be applied. This method works for the
EMCAL up to $p_t\sim 10-20$ GeV/c and for the PHOS up to $p_t\sim
30-50$ GeV/c. At higher momenta the photon clusters start to merge,
and the invariant mass analysis can not be applied.  However,
high-$p_t$ $\pi^0$'s produce non-spherical clusters which allows to
identify them via a shower shape analysis. With this method one can
distinguish $\pi^0$ and $\gamma$ in the EMCAL up to $p_t\sim 30$ GeV/c
\cite{EMCAL-TDR} and in the PHOS up to $p_t\sim 80$ GeV/c
\cite{ALICE-PPR}. For the heavier mesons, $\eta$ and $\omega$, the
$p_t$ scales are moved accordingly to the increased masses. Therefore
the main concern in the measurement of the neutral meson spectrum in
pp collisions is a limited statistics.

\begin{figure}[ht]
\centering %%
\includegraphics[width=0.45\textwidth]{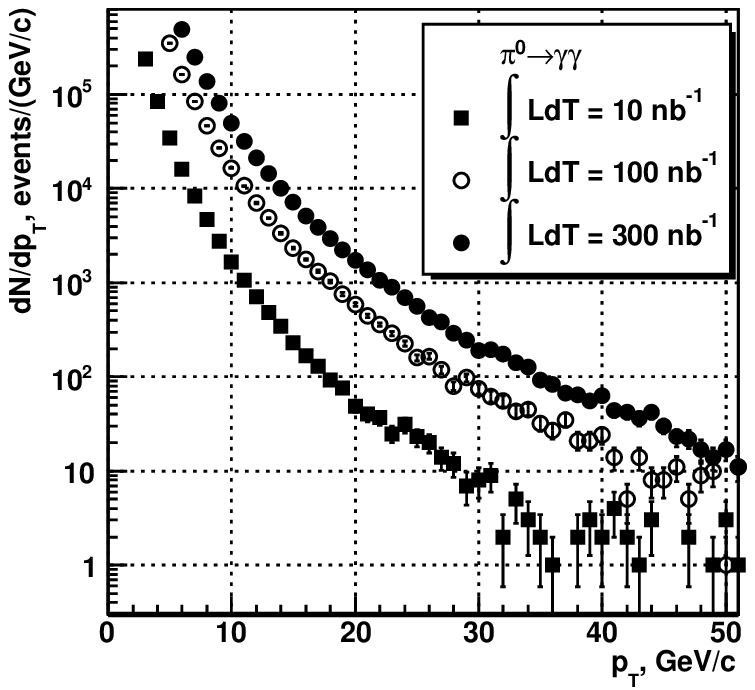}
\includegraphics[width=0.45\textwidth]{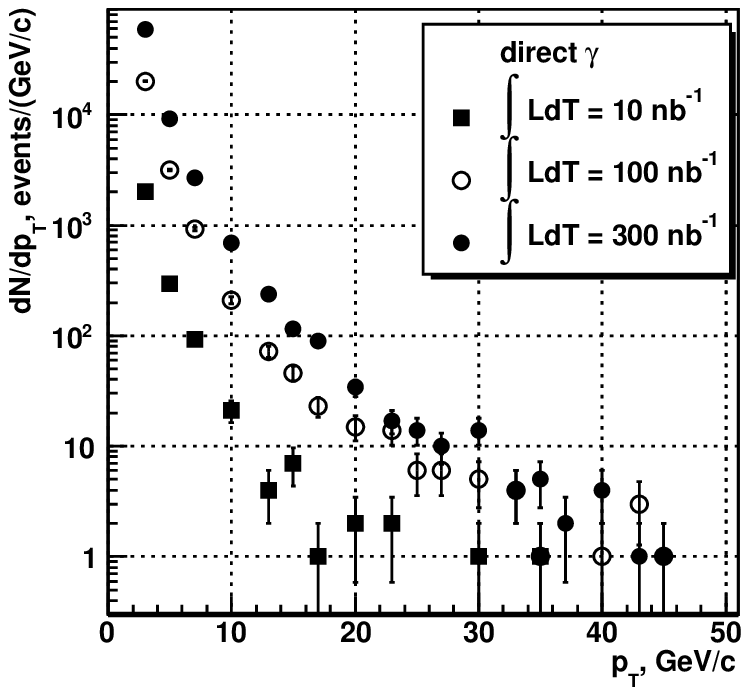}
\caption[]{The number of $\pi^0$ (left) and direct photons (right)
registered in PHOS in pp collisions at $\sqrt{s}=10$ TeV for several
values of the integrated luminosity corresponding to different
scenarios of the first LHC run.} \label{pp-range}
\end{figure}

To increase collected statistics at high $p_t$, both the PHOS and
the EMCAL define high $p_t$ triggers on energy deposited in $2\times
2$ or $4\times 4$ tiles. The transition Radiation Detector (TRD)
investigates the possibility to define a trigger on a hard converted
photon using $e^+e^-$ pair with high $p_t$ and small opening angle.
Applying these triggers we will increase the integrated luminosity,
collected in the first ALICE pp run from $\int L=10$ $nb^{-1}$
(ALICE Min.Bias) to 100 $nb^{-1}$, corresponding to the initial LHC
luminosity and even 300 $nb^{-1}$, corresponding to the optimal LHC
luminosity. Using the NLO pQCD predictions of direct photon and
$\pi^0$ yields \cite{Aurenche}, and applying PHOS acceptance and
efficiency \cite{Kharlov}, we calculated the number of $\pi^0$ and
$\gamma^{direct}$ detected per 1 GeV bin of $p_t$ for 3 values of
integrated luminosity, see fig.\ \ref{pp-range}. So, in the most
probable scenario of LHC running we can measure the $\pi^0$ (direct
photon) spectrum in PHOS up to $p_t\sim 30$ (12) GeV/c without the
PHOS trigger and 40-50 (20-30) GeV with the PHOS photon trigger.

\begin{figure}[ht]
\centering %%
\includegraphics[width=0.45\textwidth]{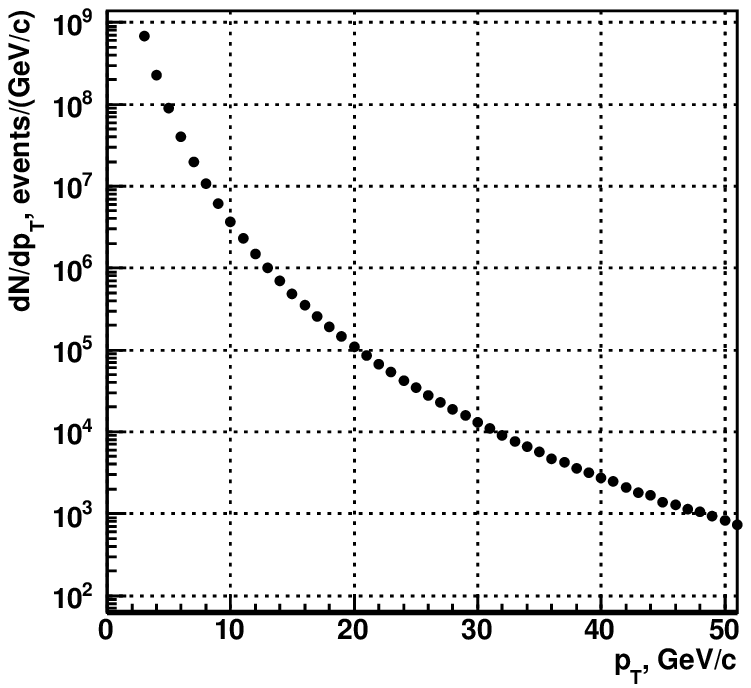}
\includegraphics[width=0.45\textwidth]{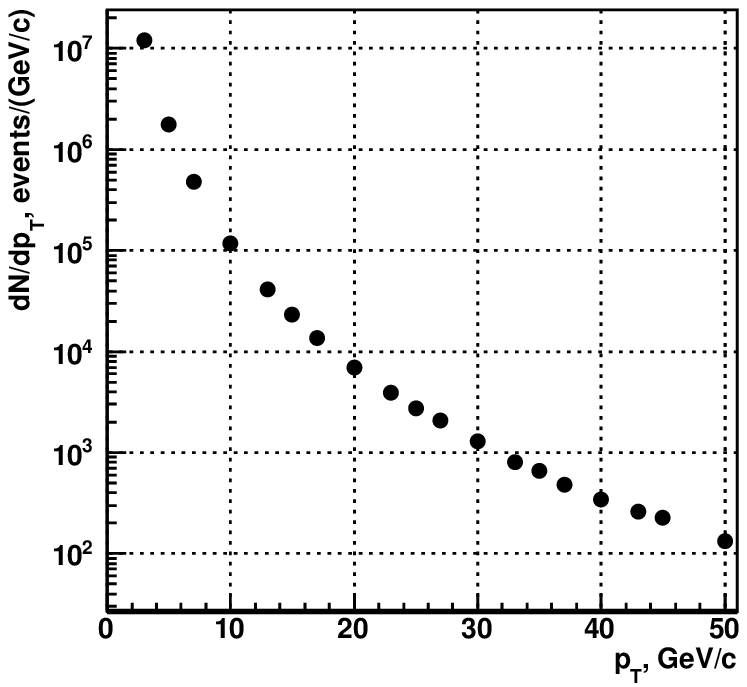}
\caption[]{The number of $\pi^0$ (left) and direct photons (right)
registered in the PHOS in 1 month run of Pb+Pb collisions at
$\sqrt{s_{NN}}=5.5$ TeV.} \label{PbPb-range}
\end{figure}

Extraction of neutral mesons in AA collisions becomes even more
complicated because of the very large combinatorial background.
However, due to the good resolutions of the PHOS detector, the
$\pi^0$ peak is clearly seen down to soft $p_t\sim 1$ GeV/c even in
central Pb+Pb collisions. Similar to the pp collisions we estimate
the accessible range of $\pi^0$ and direct photon extraction in
Pb+Pb collisions at 5.5 TeV. We used binary scaled spectra
\cite{Aurenche} in pp collisions multiplied by nuclear modification
factors: unit for photons and 0.2 for $\pi^0$, see fig.\
\ref{PbPb-range}. In this case we will be able to measure the
$\pi^0$ spectrum up to $p_t\sim 70$ GeV/c and direct photons up to
$60$ GeV/c.

\section{Gamma-jet and gamma-hadron correlations}

\begin{figure}[ht]
\centering %%
\includegraphics[width=0.45\textwidth]{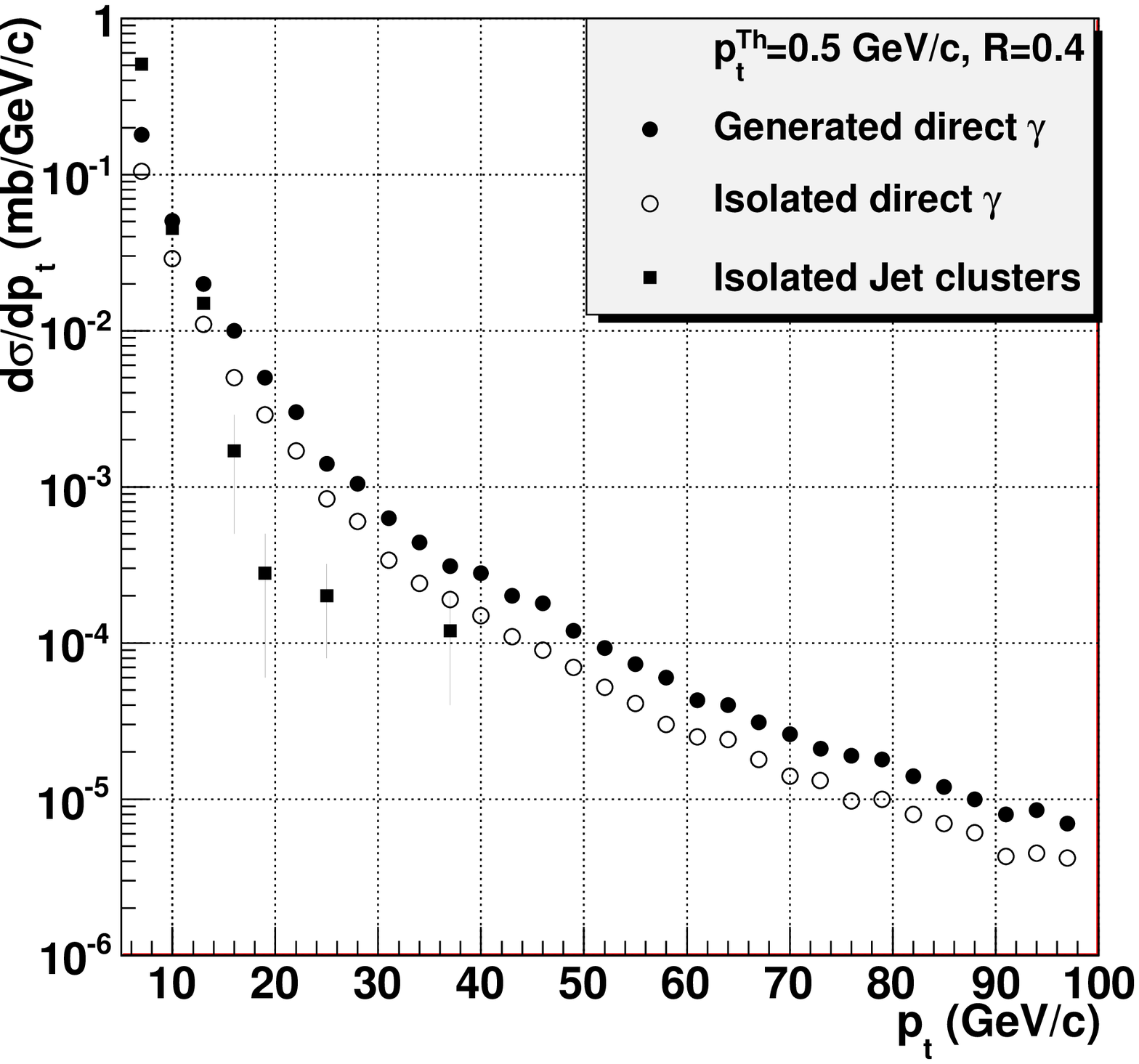}
\includegraphics[width=0.45\textwidth]{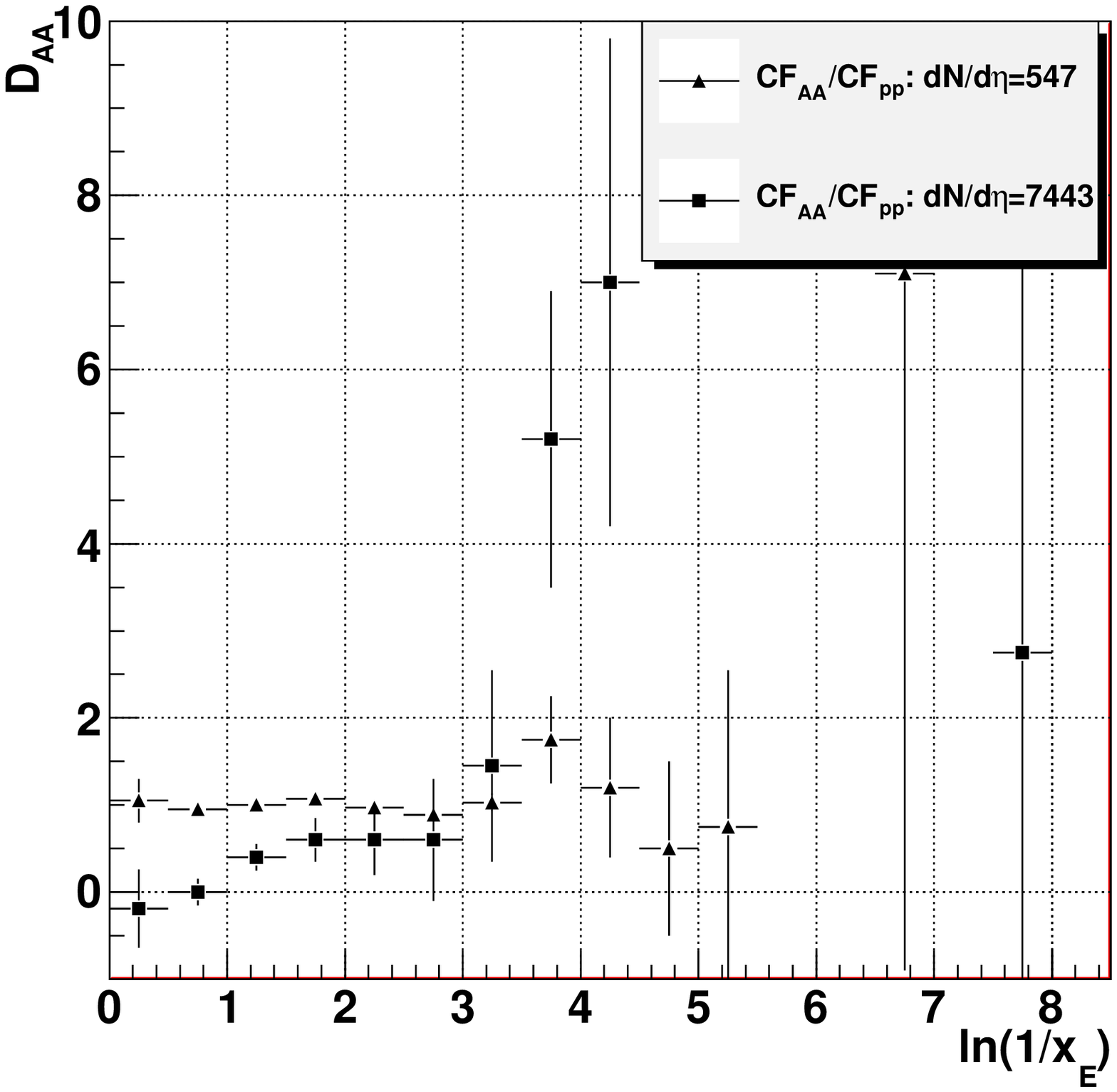}
\caption[]{Left plot: comparison of the inclusive and isolated hard
direct photon spectra with contamination from jet fragments passed
isolation cut. Right plot: modification of the imbalance functions
in Pb+Pb collisions at two values of the total multiplicity.}
\label{daa}
\end{figure}

Hard direct photons can be emitted both in hard collisions of
partons of colliding nucleons and in jet fragmentation. Direct
photons of the first type are not accompanied by hadrons and thus
can be identified in event-by-event basis using isolation technique.
Isolation is effective down to $p_t\sim 10$ GeV/c, in pp collisions
and $p_t\sim 20$ GeV/c in Pb+Pb collisions \cite{Gustavo}. To
illustrate this we present simulations of the EMCAL response to
$\gamma$-jet event in Pb+Pb collision. The signal ($\gamma$+jet) is
generated using Pythia and the underlying event with Hijing ($\hat
q=50$ $GeV^2/fm$). Spectra of all and isolated hard direct photons
as well as jet remnants which passed the isolation cut are shown in
fig.\ \ref{daa}, left plot. Contamination becomes comparable with
the signal at $p_t\sim 20$ GeV/c.

 Using the isolated photons one can construct the best possible approximation of the parton
fragmentation function --- hadron conditional yield or imbalance
function. The only difference between the true fragmentation
function and the imbalance function is the initial state radiation
($k_t$ smearing). Having constructed the imbalance functions both in
pp and AA collisions one can study the modification of the imbalance
function in AA collisions as a deviation of the ratio $D_{AA}$ from
the unit, see fig.\ \ref{daa}, right plot.

\section{Conclusions}

Photons in ALICE can be measured in two dedicated electromagnetic
calorimeters and in addition using photon conversion in the central
tracking system. All methods provide comparable resolutions but
different systematic errors. We provided an estimate of the
accessible range in measurement of $\pi^0$ and direct photon spectra
in upcoming pp and Pb+Pb LHC runs. In addition, the feasibility of
measuring isolated photon spectra, to construct the $\gamma$-jet and
$\gamma$-hadron correlations in pp and Pb+Pb collisions and studying
modification of fragmentation functions in medium, were estimated.

\section*{Acknowledgments}
We thank RFBR grant 09-02-08047-z for support.

 % do not change
\end{document}